\documentclass[preprint]{aastex}

\def\kms{\ifmmode{\rm km\thinspace s^{-1}}\else km\thinspace s$^{-1}$\fi}
\def\ms{\ifmmode{\rm m\thinspace s^{-1}}\else m\thinspace s$^{-1}$\fi}

\shorttitle{\mbox{TrES-5}}

\begin{document}

\title{\mbox{TrES-5}: A Massive Jupiter-sized Planet Transiting A Cool G-dwarf}

\author{Georgi Mandushev\altaffilmark{1}, Samuel N. Quinn\altaffilmark{2},
Lars A. Buchhave\altaffilmark{3}, Edward W. Dunham\altaffilmark{1}, 
Markus Rabus\altaffilmark{4,5}, Brian Oetiker\altaffilmark{6}, 
David W. Latham\altaffilmark{2}, David Charbonneau\altaffilmark{2}, 
Timothy M. Brown\altaffilmark{7}, Juan A. Belmonte\altaffilmark{5,8},
Francis T. O'Donovan\altaffilmark{9}}

\altaffiltext{1} {Lowell Observatory, 1400 W Mars Hill Rd, Flagstaff, AZ 
86001; gmand@lowell.edu}
\altaffiltext{2} {Harvard-Smithsonian Center for Astrophysics, 60 Garden 
St, Cambridge, MA 02138}
\altaffiltext{3} {Niels Bohr Institute, Copenhagen University, 
DK-2100 Copenhagen, Denmark}
\altaffiltext{4} {Departamento de Astonom\'ia y Astrof\'isica, Pontiﬁcia 
Universidad Cat\'olica de Chile, Casilla 306, Santiago 22, Chile}
\altaffiltext{5} {Instituto de Astrof{\'\i}sica de Canarias, C/ v{\'\i}a 
L\'actea s/n, 38200 La Laguna, Tenerife, Spain}
\altaffiltext{6} {Sam Houston State University, Huntsville, TX 77340, USA}
\altaffiltext{7} {Las Cumbres Observatory Global Telescope, 6740 Cortona Dr,
Suite 102, Goleta, CA 93117}
\altaffiltext{8} {Departamento de Astrof\'isica, Universidad de la Laguna, Spain}
\altaffiltext{9} {Ab Initio Software, 201 Spring St, Lexington MA 02421}

\begin{abstract}

We report the discovery of \mbox{TrES-5}, a massive hot Jupiter that transits 
the star GSC~03949-00967 every 1.48~days. From spectroscopy of the star we 
estimate a stellar effective temperature of $T_{\rm eff} = 5171 \pm 36$~K, and 
from high-precision $B$, $R$ and $I$ photometry of the transit we constrain the 
ratio of the semi-major axis $a$ and the stellar radius $R_\star$ to be 
$a/R_\star = 6.07 \pm 0.14$. We compare these values to model stellar 
isochrones to obtain a stellar mass of $M_\star = 0.893 \pm 0.024$~$M_\sun$. 
Based on this estimate and the photometric time series, we constrain the 
stellar radius to be $R_\star = 0.866 \pm 0.013$~$R_\sun$ and the planet radius 
to be $R_{\rm p} = 1.209 \pm 0.021$~$R_{\rm J}$. We model our radial-velocity 
data assuming a circular orbit and find a planetary mass of 
$1.778 \pm 0.063$~$M_{\rm J}$. Our radial-velocity observations rule out 
line-bisector variations that would indicate a specious detection resulting 
from a blend of an eclipsing binary system. \mbox{TrES-5} orbits one of the 
faintest stars with transiting planets found to date from the ground and 
demonstrates that precise photometry and followup spectroscopy are possible, 
albeit challenging, even for such faint stars.
\end{abstract}


\section{Introduction}

Long before the initial discoveries of extrasolar planets around Sun-like stars 
by the Doppler technique \citep{Lath89,Mayo95}, it was recognized that planets 
around other stars could also be detected photometrically if they transit their 
host stars \citep{Stru52}. Transit observations can provide at least two 
crucial pieces of information: the planet's size and orbital inclination. When 
combined with radial velocity data they can be used to derive the planet's 
mass, surface gravity and density, and infer its internal structure and 
composition. Many other studies also become possible for transiting planets 
around bright host stars, including analysis of the planet's atmosphere, 
temperature distribution, the presence of other bodies in the system, the 
spin-orbit alignment, etc. \citep[see, {\em e.g.},][]{Char07}.

The ever increasing number of discovered transiting planets, both from 
ground-based surveys and from the {\em Kepler} and {\em CoRoT} missions 
\citep{Boru10,Bagl09}, shows the tremendous diversity of extrasolar planets 
and planetary systems. Particularly interesting are the relations between 
planetary mass, radius and orbital period, as well as the distribution of 
planetary masses, as together they hold the key to understanding planetary 
formation and evolution. The current summary of transiting exoplanets at 
\verb=http://exoplanet.eu= indicates that 70\% of the transiting planets with 
known masses have masses $M_{\rm p}\lesssim 1.5 \ M_{\rm J}$. This apparent 
drop in the number of massive planets \citep[noted also by][]{Sout09} cannot 
be explained by major selection effects. Massive planets are also large, and 
thus easier to discover, unlike lower-mass planets, whose radius increases 
monotonically with mass for non-irradiated planets with cores \citep{Fort07}.

We report here the discovery of the Jupiter-sized massive transiting planet 
\mbox{TrES-5}. The planet orbits one of the faintest star yet ($V = 13.72$) 
with a transiting planet found from ground-based wide-field surveys, and is 
one of the very few planets found around stars fainter than $V = 13.5$, 
including the space-based {\em Kepler} and {\em CoRoT} missions. \mbox{TrES-5} 
increases the relatively small number of well-characterized massive transiting 
planets. Such planets play an important r\^{o}le in the study of planet 
formation, migration and evolution \citep{Chat08,Bara10} and it has been 
suggested \citep{Sout09} that the massive extrasolar planets form a different 
planet population than the lower-mass ones.

\section{Photometry and Spectroscopy\label{sec:observations}}

\begin{figure}
\caption{The discovery light curve of \mbox{TrES-5}. The plot shows the 
relative flux of the TrE5-5 system as a function of the orbital phase, 
adopting the ephemeris in Table~\ref{t:planet}. \label{f:discovery}}
\plotone{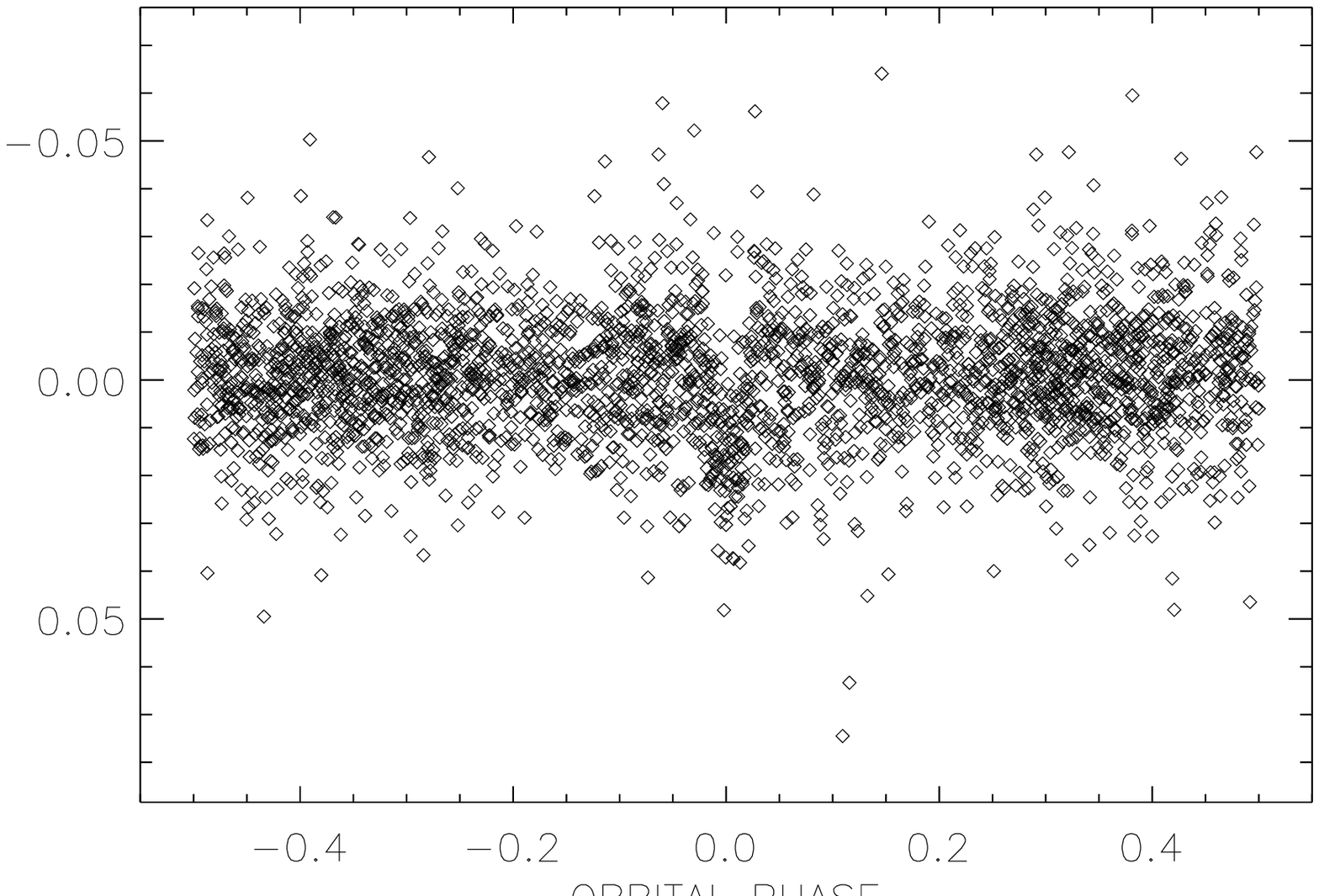}
\end{figure}

We monitored a $5\fdg8 \times 5\fdg8$ field in Cygnus with the Lowell 
Observatory Planet Search Survey Telescope \citep[PSST, ][]{Dunh04} and the 
STARE telescope on the Canary Islands in Spain, between UT 2007 July 15 and 
UT 2007 October 7. Both telescopes are part of TrES (the other two telescopes 
--- Sleuth, at Palomar Observatory, and WATTS, in Texas, were not operational 
at that time). The same field was observed again in 2009 and 2010 by PSST 
alone. All images were reduced, and photometry and transit search carried out 
as described in \citet{Dunh04}. Altogether PSST observed 13 full and partial 
transit-like events of the star GSC~03949-00967. The object was at the very 
edge of the PSST field and because of the slightly different field size and 
scale the star was not observed by the STARE telescope. The depth and duration 
of the events were consistent with the transit of a Jupiter-sized planet across 
a K dwarf, and we undertook a program of followup observations to confirm the 
planetary nature of the object and measure its properties. The discovery light 
curve of TrES-5 is shown on Fig.~\ref{f:discovery}

\begin{figure}
\caption{High-precision followup $B$, $V$ and $R$-band photometry of 
\mbox{TrES-5}. The plot shows the relative flux of the TrE5-5 system as a 
function of time relative to the center of transit, adopting the ephemeris in 
Table~\ref{t:planet}. Each light curve is labeled with the telescope and 
date of observation. The residuals from the simultaneous fits (overplotted 
with solid lines) are shown below each light curve. \label{f:followup}}
\epsscale{0.95}
\plotone{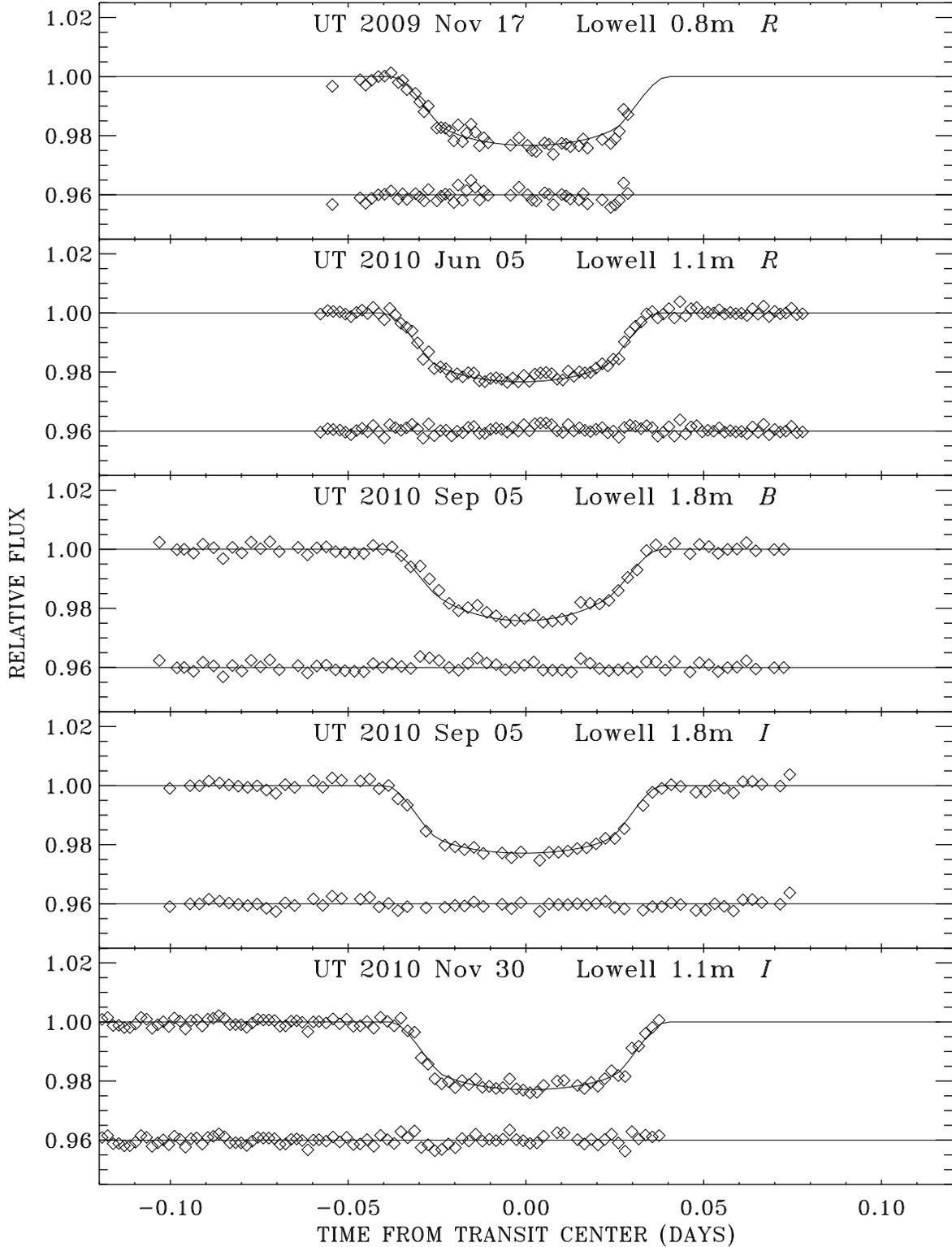}
\end{figure}

We carried out high-precision in-transit $B$, $R$, and $I$-band photometry of 
\mbox{TrES-5} using Lowell Observatory's 0.8m, 1.1m Hall and 1.8m Perkins 
telescopes. We observed \mbox{TrES-5} on UT 2009 November 17 with the 0.8m 
telescope with NASACam in Cousins $R$ (47 exposures), on UT 2010 June 5 and 
2010 November 30 with the 1.1m Hall telescope with the NASA42 camera in Cousins 
$R$ and $I$ (87 and 108 exposures, respectively), and on UT 2010 September 5 
with the 1.8m Perkins telescope with PRISM in Johnson $B$ and Cousins $I$ 
(64 and 57 exposures, respectively). For all data sets, we derived differential 
fluxes relative to a large ensemble of local comparison stars. The 
high-precision Lowell Observatory photometry is shown in Fig.~\ref{f:followup}.

In order to characterize the host star and provide an external check on the 
stellar parameters derived from the TRES spectra, we obtained off-transit 
$BV(RI)_{\rm C}$ photometry of the \mbox{TrES-5} system on UT 2010 October 10 
and 11 with the 1.05-m Hall telescope at Lowell Observatory in combination with 
a $4{\rm K} \times 4{\rm K}$ e2v CCD231 CCD detector. We calibrated the 
photometry using 6 standard fields from \citet{Land92}. The results are listed 
in Table~\ref{t:host_star} together with other relevant data for the host star 
of \mbox{TrES-5}.

We observed the candidate with the Tillinghast Reflector Echelle Spectrograph 
\citep[TRES;][]{Fure08}, mounted on the 60-inch Tillinghast Reflector at 
Fred L. Whipple Observatory on Mt. Hopkins in Arizona from 2010 September to 
2011 April. We obtained 8 spectra, each consisting of 51 echelle orders and 
spanning the wavelength range 3850-9100~\AA, with a resolving power of 
$\lambda/\Delta\lambda \approx 44,\!000$. We obtained absolute radial 
velocities by cross-correlation against a synthetic template chosen from a 
large library of spectra based on Kurucz model atmospheres 
\citep[see][]{Nord94, Lath02}, calculated by John Laird and based on a linelist 
compiled by Jon Morse. These velocities have a typical precision of 0.2~\kms. 
However, the synthetic spectra only span a small portion of the total 
wavelength range, centered on the Mg~b triplet at $5187$~\AA. We derived 
precise relative velocities by taking advantage of the full spectrum, 
cross-correlating all orders against those of an observed template. For a 
description of the reduction and cross-correlation, see \cite{Buch10}, in which 
the details of the same reduction package are described, as applied to the 
Fiber-fed Echelle Spectrograph (FIES). The absolute gamma velocity, calculated 
using the weighted mean offset of the relative velocities from the 
corresponding absolute velocities, was determined to be 
$\gamma = -13.403 \pm 0.100$~\kms. The absolute gamma velocity is onto the 
system in which HD 182488 is defined to have an absolute velocity of 
$-21.508$~\kms.

To derive the stellar atmospheric parameters, the observed spectra were 
cross-correlated against a grid of synthetic spectra drawn from the library of 
spectra described above. The synthetic spectra cover a window of $300$~\AA \ 
centered near the gravity-sensitive Mg~b triplet and have a spacing of $250$~K 
in effective temperature, 0.5~dex in gravity, 0.5~dex in metallicity and 
$1$~\kms \ in rotational velocity. The best matched template to the observed 
spectrum represents the best matched stellar parameters on the library grid. A 
new set of tools is then used to derive more precise stellar parameters from 
the normalized cross correlation peaks. A description of the tools will be 
published in Buchhave et al. (2011, in preparation).

We used the multicolor Johnson-Cousins photometry to form a variety of color 
indices, and used the color-temperature calibrations by \cite{Rami05} and 
\cite{Casa06} to estimate the star's effective temperature $T_{\rm eff}$. The 
average result, $T_{\rm eff} = 4943 \pm 36$~K, is 228~K cooler than the 
spectroscopic value of $T_{\rm eff} = 5171 \pm 36$~K (see below). This result 
is not surprising as some reddening is expected at the object's Galactic 
latitude of $13^{\circ}$. From the match to stellar evolution models described 
below, we estimate a reddening of $E_{B-V} \approx 0.07$, which is in good 
agreement with the difference of 228~K between the spectroscopic and 
photometric effective temperatures.

\section{Light Curve Analysis\label{sec:lightcurve}}

We analyzed the five $B$, $R$ and $I$-band transit observations using the 
analytical expressions in \citet{Mand02} to compute the model flux. We 
assumed a circular orbit and a quadratic stellar limb-darkening law, fixing the 
coefficients at the color-dependent values tabulated in \citet{Clar04} for the 
spectroscopically-estimated $T_{\rm eff}$, $\log g$, and  
$\left [ {\rm Fe/H} \right ]$. Initially we adoped as free parameters the 
orbital period $P$ and epoch $T_0$, the ratio of the planet radius to the 
stellar radius $R_{\rm p}/R_\star$, the square of the impact paramater 
$b = (a/R_\star)\cos i$, where $a$ is the semi-major axis of the planet's 
orbit and $i$ is the orbital inclination, the parameter $\zeta /R_\star$, 
which for zero eccentricity is related to $a/R_\star$ via 
$\zeta /R_\star = (a/R_\star)(2\pi/P)(1-b^2)^{-1/2}$, and the zero point of the 
out-of-transit flux $F_0$. This choice of parameters minimizes the correlation 
between $b$ and $a/R_\star$ \citep[see][]{Bako07}. We performed an initial fit 
using the high-precision followup transit photometry, after which we fixed all 
parameters except $P$ and $T_0$ to their values from the initial fit, and 
repeated the fit using both the PSST wide-field photometry and the followup 
photometry. This approach allowed us to utilize the entire baseline of over 
three years of observations and obtain more precise values for $P$ and $T_0$ 
than if only the followup photometry were used. We then fixed the values of 
$T_0$ and $P$ (stated in Table~\ref{t:planet}) in the subsequent analysis.

We performed a joint fit of all five transit light curves using the MPFIT 
package \citep{Mark08}, and found the values of $R_{\rm p}/R_\star$, $b^2$, 
$\zeta /R_\star$ and $F_0$ that minimized the ${\chi}^{2}$. This fit is shown 
with solid lines in Fig.~\ref{f:followup}. In order to estimate the errors of 
the fitted parameters, we conducted a Markov Chain Monte Carlo (MCMC) analysis 
\citep[see][and references therein]{Ford05}. Altogether 10 independent Markov 
chains with $5\times10^5$ points per chain were created, each chain starting 
from a random perturbation of the best-fit parameters. We discarded the first 
20\% of the points to minimize the impact of the initial conditions. The 
resulting histograms of $4 \times 10^6$ values were used to obtain the mode and 
the 68.3\% lower and upper confidence limits $p_{\rm lo}$ and $p_{\rm up}$ for 
each parameter. We then adopted the mode as the final, optimal value of the 
parameter, and the value of $(p_{\rm up} - p_{\rm lo})/2$ as its 1-$\sigma$ 
error (encompassing 68.3\% of the parameter values around the mode). The use of 
a single number to characterize the width of each parameter's distribution is 
justified by the nearly Gaussian, symmetric shape of all histograms.

\section{Properties of \mbox{TrES-5} And Its Star\label{sec:star}}

The mass and radius of the host star of \mbox{TrES-5}, required for 
establishing the planet properties, were determined on the basis of the 
spectroscopic $T_{\rm eff}$ and $\left [ {\rm Fe/H} \right ]$, and the value of 
$a/R_\star$ derived from the light curve fit described above. The quantity 
$a/R_\star$ is closely related to the stellar density, and is determined in 
this case with higher relative precision than $\log g$. It is therefore a 
better proxy for luminosity \citep[see][]{Sozz07}. 

We used an iterative procedure to determine the star's parameters, similar 
to the one described in \cite{Torr08}. Using the stellar evolution models from 
the series by \cite{Yi01}, we computed isochrones over a range in stellar age, 
from 0.1~Gy to 12~Gy, over the metallicity range allowed by the spectroscopy. 
We compared the measured $T_{\rm eff}$ and $a/R_\star$ from the light curve fit 
to the computed values of $a/R_\star$ and $T_{\rm eff}$ along each isochrone 
and recorded the points on the isochrone which matched the measured quantities 
within their errors. We then calculated the weighted mean of all matches, with 
weights inversely proportional to the exponent of the distance (in $\chi^2$ 
sense) between the observed and model values of $T_{\rm eff}$, $a/R_\star$ and 
$\left [ {\rm Fe/H} \right ]$ \citep[see][]{Torr08}.

This procedure yields a value of $\log g$ which is better constrained than the 
spectroscopic estimate. With the value of $\log g$ fixed, revised estimates of 
$T_{\rm eff}$ and $\left [ {\rm Fe/H} \right ]$ were derived from the spectra, 
and the light curve fit repeated. The new values of $T_{\rm eff}$ and 
$\left [ {\rm Fe/H} \right ]$ (from spectroscopy), and $a/R_\star$ (from the 
fit) were again compared to the stellar models to obtain the final values of 
$M_\star = 0.893 \pm 0.024 \ M_\sun$, $R_\star = 0.866 \pm 0.013 \ R_\sun$, 
$\log g = 4.513 \pm 0.013$, and age of $7.38 \pm 1.87$~Gy.

\begin{figure}
\caption{{\em Top}: Radial velocity observations of \mbox{TrES-5} obtained with 
TRES, shown relative to the center of mass and adopting the ephemeris in 
Table~\ref{t:planet}. The best-fit orbit ({\em solid line}) is overplotted. 
{\em Middle}: Residuals from the best-fit model to the radial velocities. 
{\em Bottom}: Bisector spans shifted to a median of zero, for each of the 
TRES exposures. \label{f:rv}}
\plotone{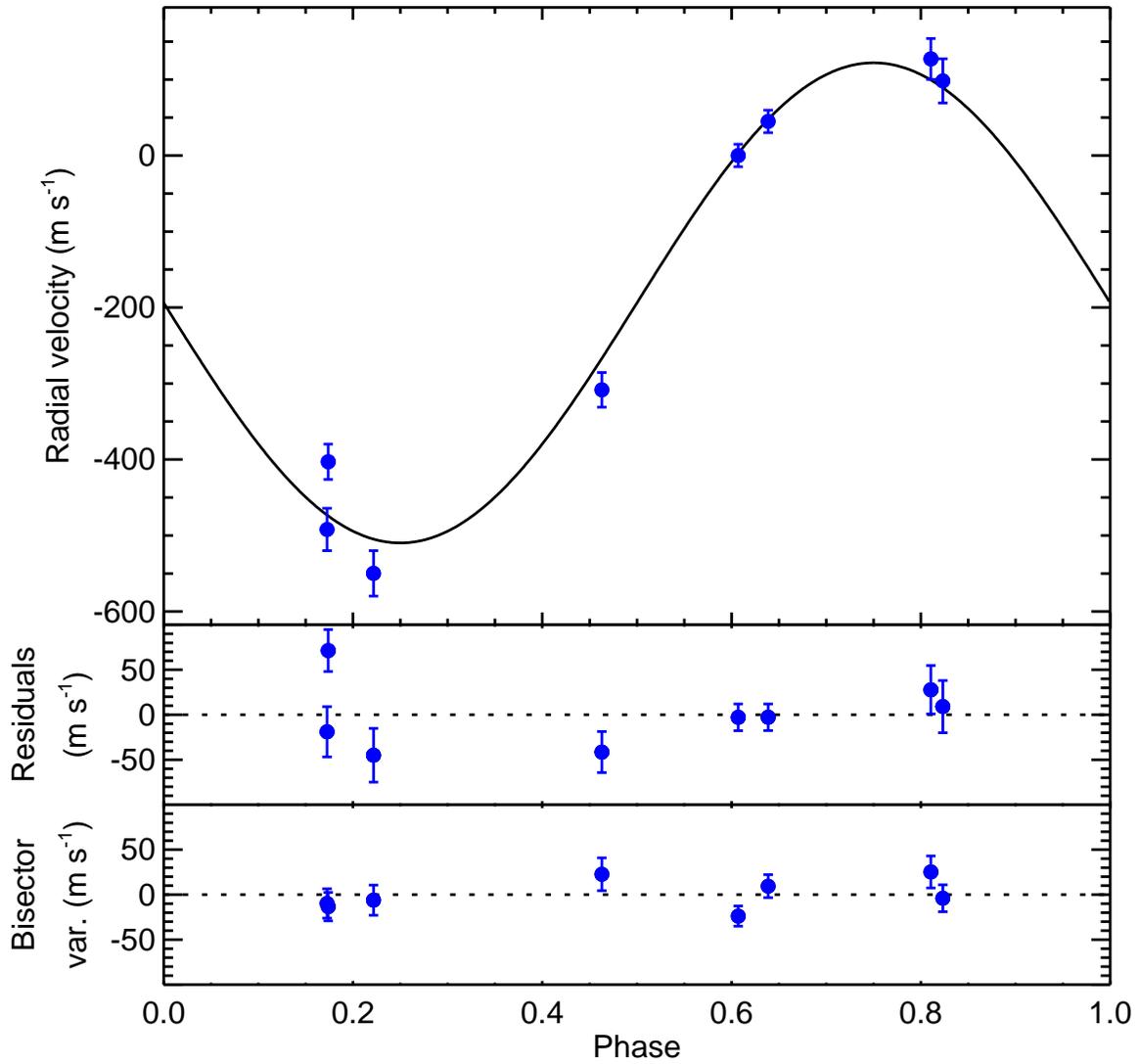}
\end{figure}

We fit a Keplerian orbit to these data assuming zero eccentricity as a good 
first approximation, as expected from theoretical arguments for a period as 
short as 1.48~days. The period and epoch were held fixed. The RMS of this fit 
is 24.2~\ms, which is similar to the internal errors of the velocities. The 
parameters of this orbital solution are listed in Table~\ref{t:planet}. The 
orbit is displayed in Fig.~\ref{f:rv} (top panel) along with the observations, 
and the residuals are shown in the middle panel.

We investigated the possibility that the radial velocities we measured are the 
result of distortions in the line profiles due to contamination from an 
unresolved eclipsing binary \citep{Sant02,Torr05} or star spots \citep{Tone88}, 
instead of being due to true Doppler motion in response to a planetary 
companion. We cross-correlated each TRES spectrum against a synthetic template 
matching the properties of the star, and averaged the correlation functions 
over the orders. From this representation of the average spectral line profile 
we computed the mean bisectors, and as a measure of the  line asymmetry we 
calculated the ``bisector spans" as the velocity difference  between points 
selected near the top and bottom of the mean bisectors \citep{Tone88,Torr05}. 
If the velocities were the result of a blend with an eclipsing binary 
\citep{Mand05} or star spots \citep{Quel01}, we would expect the line bisectors 
to vary in phase with the photometric period with an amplitude similar to that 
of the velocities. Instead, we detect no variation in excess of the measurement 
uncertainties (see Fig.~\ref{f:rv}, bottom panel), and we conclude that the 
velocity variations are real and that the star is orbited by a Jovian  planet.

\section{Discussion}

\mbox{TrES-5} orbits one of the faintest host stars with transiting planets 
found to date. The discovery demonstrates that precise photometry and 
spectroscopy are possible from the ground even for such faint stars, although 
some followup studies will be more challenging. With a period of only 
$1.5$~days, \mbox{TrES-5} is a classical ``hot Jupiter'', although the 
relatively cool host star ($T_{\rm eff} = 5171$~K) places it near the upper 
(hotter) end of the transition zone between the ``pM class'' and ``pL class'' 
planets proposed in \cite{Fort08}. In such planets the importance of gaseous 
TiO and VO absorption is diminished, temperature inversion may not be present, 
and the flux from the host star is more evenly distributed around the planet 
and inside its interior. Because of its location near the pM/pL boundary, 
\mbox{TrES-5} could be a suitable object for testing the models of planetary 
atmospheres. 

At $M_{\rm p} = 1.8 \ M_{\rm J}$ \mbox{TrES-5} is above the upper quartile of 
the planetary mass distribution. The mass and radius of \mbox{TrES-5} are in 
good agreement with the theoretical mass-radius relations for irradiated 
planets \citep{Fort07,Hans07}. For an age of 7.5~Gyr, planetary mass of 
$M_{\rm p} = 1.8 \ M_{\rm J}$ and equilibrium temperature of 
$T_{\rm eq} \sim 1500$~K, the predicted radius of the planet is 
$R_{\rm p} \sim 1.2 \ R_{\rm J}$, which is in good agreement with the observed 
value. Thus, \mbox{TrES-5} does not appear to have the anomalously large radius 
of many close-in giant exoplanets. A comparison with the planetary models of 
\cite{Fort07} for the mass, age and semimajor axis of \mbox{TrES-5} indicates 
that the planet is best approximated by a model with no heavy-elements core.

It has been noted that many massive planets have non-zero eccentricity even at 
short periods \citep{Sout09}. Using the current catalog of transiting planets 
at \verb=http://exoplanet.eu=, we find that over a third of the short-period, 
massive planets ($P < 3$~days and $M_{\rm p} > 1.7 \ M_{\rm J}$) have non-zero 
eccentricities. In our radial velocity fit we have assumed $e = 0$ based on the 
short period of \mbox{TrES-5} and theoretical arguments about the time scales of 
tidal interactions \citep{Zahn89}. Our radial velocities clearly reject large 
eccentricities. From MCMC analysis of the radial velocity data, we find that 
the prefered orbital solution has $e = 0.025^{+0.015}_{-0.025}$. We therefore 
adopt $e = 0$ for the orbit of \mbox{TrES-5}, but we cannot rule out a small 
eccentricity ($e < 0.04$). A future detection of the secondary eclipse will 
help constrain the orbital eccentricity much better, as well as test for the 
presence of temperature inversion in the atmosphere of \mbox{TrES-5}.

\acknowledgments

We thank Travis Barman for useful discussions. This paper is based on work 
supported in part by NASA grants NNG04GN74G, NNG04LG89G, NNG05GI57G, 
NNG05GJ29G, and NNH05AB88I through the Origins of Solar Systems Program, and 
NASA Planetary Major Equipment grant N4G5-12229. We acknowledge support from 
the NASA {\em Kepler} mission under cooperative agreement NCC2-1390, and 
M.~R.\ acknowledges support from ALMA-CONICYT projects 31090015 and 31080021. 
This publication makes use of data products from the Two Micron All Sky Survey, 
which is a joint project of the University of Massachusetts and the Infrared 
Processing and Analysis Center/California Institute of Technology, funded by 
NASA and the National Science Foundation.

\clearpage

\clearpage

\clearpage

\begin{deluxetable}{llcc}
\tablewidth{0pt}
\tablecaption{\mbox{TrES-5} Host Star \label{t:host_star}}
\tablehead{
\colhead{Parameter} & \colhead{Units} & \colhead{Value} & \colhead{Source}}
\startdata
RA                            & J2000.0       & $20^{\rm h} 20^{\rm m} 53\fs 24$  & 1 \\
Decl.                         & J2000.0       & $+59\arcdeg 26\arcmin 55\farcs 6$ & 1 \\
GSC                           &               & 03949-00967                       &   \\
$[\mu_{\alpha},\mu_{\delta}]$ & mas~yr$^{-1}$ & $\left [+10.4,+30.5 \right ]$     & 1 \\
$V$                           &               &     $13.718 \pm 0.005$            & 2 \\
$B-V$                         &               & \phn$ 0.927 \pm 0.008$            & 2 \\
$V-R_{\rm C}$                 &               & \phn$ 0.526 \pm 0.007$            & 2 \\
$V-I_{\rm C}$                 &               & \phn$ 0.994 \pm 0.007$            & 2 \\
$J$                           &               &     $12.111 \pm 0.027$            & 3 \\
$J-H$                         &               & \phn$ 0.438 \pm 0.035$            & 3 \\
$J-K_s$                       &               & \phn$ 0.547 \pm 0.033$            & 3 \\
$M_\star$                     & $M_\sun$      & \phn$ 0.893 \pm 0.024$            & 2 \\
$R_\star$                     & $R_\sun$      & \phn$ 0.866 \pm 0.013$            & 2 \\
$T_{\rm eff}$                 & K             &       $5171 \pm 36$               & 2 \\
$\left [ {\rm Fe/H} \right ]$ &               &     $+0.20  \pm 0.08$             & 2 \\
$\log g$                      &               & \phn$ 4.513 \pm 0.013$            & 2 \\
$v \sin i$                    & \kms          & \phn$ 3.80  \pm 0.36$             & 2 \\
$M_V$                         &               &     $+5.756 \pm 0.041$            & 2 \\      
Distance                      & pc            &      $360   \pm 11$               & 2 \\
Age                           & Gy            &      $7.38  \pm 1.87$             & 2 \\
\enddata
\tablerefs{(1) UCAC3 \citep{Zach10}; (2) this paper; (3) 2MASS \citep{Skru06}}
\end{deluxetable}

\begin{deluxetable}{lr}
\tablewidth{0pt}
\tablecaption{Radial velocity measurements of \mbox{TrES-5} \label{t:rv}}
\tablehead{
\colhead{HJD} & \colhead{\ \ RV (\ms)}}
\startdata
2455456.847578 &  $-542.8 \pm 27.9$ \\
2455468.707361 &  $-472.6 \pm 23.3$ \\
2455647.005138 &  $-308.3 \pm 22.8$ \\
2455649.002656 &  \phm{-}$127.2 \pm 26.9$ \\
2455651.985839 &  \phm{-} \phn $98.3 \pm 29.0$ \\
2455659.987762 &  $-549.8 \pm 29.9$ \\
2455667.970142 &  \phm{-} \phn \phn $0.0 \pm 14.8$ \\
2455673.945954 &  \phm{-} \phn $45.0 \pm 14.8$ \\
\enddata
\end{deluxetable}

\begin{deluxetable}{llr@{$\: \pm \:$}l}
\tablewidth{0pt}
\tablecaption{\mbox{TrES-5} Planet Parameters \label{t:planet}}
\tablehead{
\colhead{Parameter} & \colhead{Units} & \multicolumn{2}{c}{Value}}
\startdata
$P$                           & days          & $1.4822446$ & $0.0000007$    \\
$T_0$                         & HJD           & $2\, 455\, 443.25153$ & $0.0001069$ \\
$a$                           & AU            & $0.02446$ & $0.00068$        \\
$i$                           & deg           & $84.529$  & $0.005$          \\
$a/R_\star$                   &               & $6.074$   & $0.143$          \\
$b = a \cos i/R_\star$        &               & $0.579$   & $0.026$          \\
$K$                           & \ms           & $339.8$   & $10.4$           \\
$M_{\rm p}$                   & $M_{\rm J}$   & $1.778$   & $0.063$          \\
$R_{\rm p}$                   & $R_{\rm J}$   & $1.209$   & $0.021$          \\
$\bar \rho$ & ${\rm g \: cm^{-3}}$            & $1.25$    & $0.08$           \\
$T_{\rm eq}$                  & K             & $1484$    & $41$             \\
$R_{\rm p}/R_\star$           &               & $0.1436$  & $0.0012$         \\
\enddata
\tablecomments{$M_{\rm J} = 1.899 \times 10^{27}$~kg (Jupiter's mass); 
 $R_{\rm J} = 7.1492 \times 10^7$~m (Jupiter's equatorial radius) \citep{Cox00}}
\end{deluxetable}

\end{document}